\documentclass[12pt]{article}
\usepackage{graphicx}
\begin{document}
{\it AIP Conference Proceedings on the Monte Carlo method in the physical 
sciences, 

to be edited by J.E. Gubernatis.}

\bigskip
\centerline{\bf How to convince others ? }

\centerline{Monte Carlo simulations of the Sznajd model}

\bigskip
Dietrich Stauffer

Institute for Theoretical Physics, Cologne University, D-50923 K\"oln, 
Euroland

\begin{abstract}
In the Sznajd model of 2000, a pair of neighbouring agents on a square lattice
convinces its six neighbours of the pair opinion if and only if the two agents 
of the pair share the same opinion. It differs from other consensus models 
of sociophysics (Deffuant et al., Hegselmann and Krause) by having integer 
opinions like $\pm 1$ instead of continuous opinions. The basic results and 
the progress since the last review are summarized here.
\end{abstract}

\section{Introduction}
The application of cellular automata, Ising models and other tools of 
(computational or statistical) physics has a long tradition 
\cite{Majorana,Schelling,Callen,Galam,Schweitzer,Weidlich}. Of course, thinking
human beings are not enthusiastic about being treated like a randomly flipping
magnetic moment, since they form their opinions by complicated cognitive 
processes. But to see general properties of mass psychology, such simple 
approximations may be realistic enough. Similarly, conceiving a child is a
very private affair; nevertheless from average birth rates one can predict 
reasonably well how many babies will be born next year in a large country. 
Whether I smoke, drink or eat steaks, has influence on my health and on my 
time of death; nevertheless, average mortality rates are published in many
countries and were already studied three centuries ago by comet researcher 
Halley.
 
\begin{figure}
  \includegraphics[angle=-90,scale=0.6]{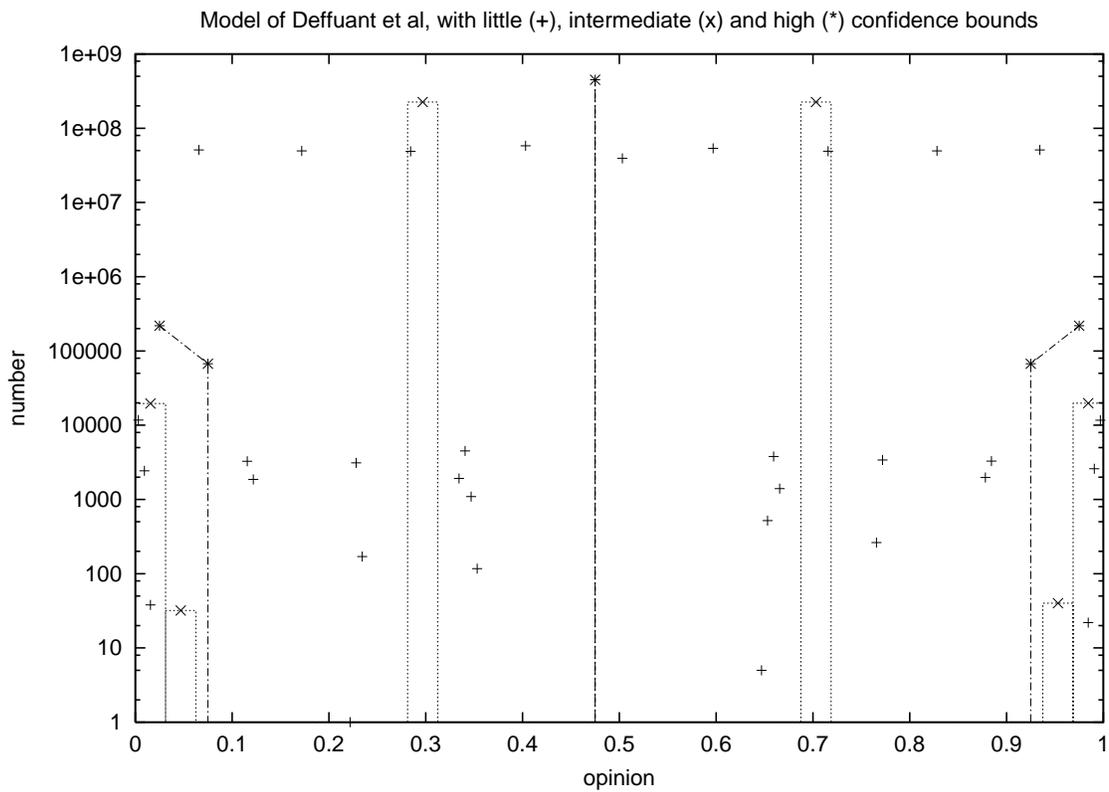}
\caption{
Can the European Union ever agree on a joint foreign policy ? The final 
distribution of opinions in the model of Deffuant et al is shown for little
tolerance against different opinions (+, fragmentation), intermediate tolerance 
($\times$, two opposing camps) and high tolerance ($\ast$, consensus). 
Note that some extremist opinions always remain.}
\end{figure}

More recently, starting perhaps with \cite{Axelrod} (see \cite{Klemm}  for 
recent simulations and further literature), the ``consensus`` literature tried
to find out when (in a computer simulation) a complete consensus from initially 
diverging opinions emerges. Deffuant et al \cite{Deffuant} (henceforth denoted
as D, where earlier papers on this model are cited) and Hegselmann and Krause 
(HK) \cite{Hegselmann} had opinions on a continuous scale between $-1$ and 1, 
while the Sznajd model \cite{Sznajd} (for a review see \cite{Stauffer}) mostly 
used the binary choice $\pm 1$ for opinions. 

The HK model and most other ``voter'' models \cite{voter} assume that every 
agent is influenced by its neighbours or by all other agents and takes, for 
example, the opinion of the majority of them, or of a weighted average, as its 
own next opinion. The Sznajd model,
on the other hand, assumes that every agent tries to influence its neighbours,
without caring about what they think first. Thus in the Sznajd model
the information flows outward to the neighbourhood, as in infection or rumour 
spreading, while in most other models the information flows 
inward from the neighbourhood. The D model is in between:
Two people who exchange opinions move closer together in their opinions. 
The models of HK and D assume that only people whose
opinions are already close to each other can influence each other: bounded
confidence. The results of these two models are similar but the D model  
is faster to simulate, up to 450 million agents by the present
author, Fig.1. The main result is that a complete consensus is reached if the 
interval of opinions over which people influence each other is large enough, 
while for small such intervals at the end several distinct opinions survive. 
For the D model this can also be seen from approximate analytical 
treatment \cite{Ben-Naim}.
  
The Sznajd model, with only the opinions $\pm 1$ allowed, always leads to a 
complete consensus, and this remains true if $Q > 2$ different opinions are 
allowed and all opinions can influence each other. With bounded confidence
in the sense that opinion $q$ can influence only the neighbouring opinions
$q \pm 1$, the results are similar to those of D and HK.
Full consensus for $Q$ up to three but not for larger $Q$. However,
the Sznajd model takes into account the
well-known psychological and political fact that ``united we stand, divided we
fall''; only groups of people having the same opinion, not divided groups, can
influence their neighbours. 

In contrast to the other consensus models, the Sznajd model as published
thus far deals only with communication between
neighbours, not between everybody. It is a ``word-of-mouth'' model.

\section{Models}

To see if a consensus emerges out of initially different opinions, all three 
models here start with a random initial distribution of opinions $S_i, \; i=1,2,
\dots N$ of $N$ ``people'', where $S_i$ is a real number (between 0 and 1) in 
the D \cite{Deffuant} and HK \cite{Hegselmann} models, while it is an integer in
the Galam \cite{galam} and the Sznajd model \cite{Sznajd}. Only the basic D and
HK versions are reviewed  here. 
Fortran programs are listed in \cite{Stauffer} or given in the appendix. 

\subsection{Deffuant et al}

In the D model \cite{Deffuant}, at every Monte Carlo step a randomly selected 
pair $i,k$ checks if the opinions $S_i$ and $S_k$ differ by less than a fixed 
parameter $\epsilon$. If no, nothing happens; if yes, both opinions move closer 
to each other by an amount $\mu |S_i-S_k|$. In Fig.1, the weight $\mu$ was taken
as 0.3, and $\epsilon = 0.05, 0.25$ and 0.40 for low, intermediate and high
tolerance of dissent.

\subsection{Hegselmann-Krause}

Also in the HK model \cite{Hegselmann}, the opinions vary between 0 and 1. 
At each Monte Carlo step, one randomly selected $i$ takes the average opinion
of all other opinions $S_k$ which differ less than $\epsilon$ from $S_i$. 
Because of this large sum, the simulation for large numbers $N$ of people is 
much slower than in the D model; nevertheless the final results are quite 
similar, Fig. 2 of \cite{sociocise}.
 
\subsection{Galam}

The Galam model \cite{galam} is not really a consensus model since dissenters
are ignored, not convinced. $N = 4^n$  people are divided into $N/4$ groups 
of four each, each group determining by majority vote which of two possible
opinions the single delegate of that group will support. Four such delegates 
again select one representative by majority vote, four such representatives
select one council member, and so on, until after $n$ such steps of majority 
hierarchies one opinion represents the whole community. In case of a 2:2 tie,
the status quo is preserved, i.e. the government wins over the opposition. 
Even if in the initial random distribution of opinions, the opposition has a
sizeable majority, at the end the minority government wins, also in the case 
of more realistic Monte Carlo simulations or modified models \cite{others}. 

\subsection{Panic}

Some sort of consensus is also reached if all people in a room on fire run to
one of two exits, leaving the other exit unused. This panic is the limiting
case of a simulation \cite{Helbing} using molecular dynamics techniques, where
in general each person follows partly the majority direction and partly his/her
own judgement.

\subsection{Sznajd}

The Sznajd people usually sit on lattice sites, and a pair of two neighbours 
$i,k$ having the same opinion $S_i = S_k$ convinces all its neighbours of this 
opinion $S_i$. Instead of a pair, also a single site, or a plaquette of
four agreeing neighbours has been simulated \cite{Ochrombel,Sousa} to convince
all neighbours. 

\section{Basic Sznajd Results}

The basic Sznajd model with random sequential updating always leads to a 
consensus, even if more than two opinions are allowed or for higher dimensions. 
If initially half of the opinions are $+1$ and the other half $-1$,
then at the end half of the samples will have $S_i=+1$ for all $i$, and the 
remaining half have $S_i=-1$ everywhere. A phase transition is often observed as
a function of the initial concentration $p$ of up spins $S_i=1$: For $p < 1/2$ 
all samples end up with $S_i=-1$, and for $p > 1/2$ they all end up in the 
other fixed point $S_i = +1$, for large enough lattices. This phase transition
at $p_c = 1/2$ does not exist in one dimension \cite{Sznajd} or when a single
site (instead of a pair or plaquette) on the square lattice \cite{Ochrombel} 
already convinces its neighbours. Pictures and cluster analysis of the
domain formation process \cite{Bernardes,Stauffer} show strong similarity with
Ising models. The time needed to reach a complete consensus fluctuates widely 
and (in the cases were a phase transition is found) does not follow a Gaussian 
or log-normal distribution. If convincing happens only with a certain 
probability, then no complete consensus is found \cite{Sznajd,Sousa}.
A Hamiltonian-like description seems possible (only ?) in one dimension
\cite{sznajdweron}. The number of people who never changed their 
opinion first decays with a power of time, and then stays at a small but finite
value \cite{pmco}, quite different from Ising models. 
See \cite{weron} for an economic application.

\section{Sznajd Modifications}

Switching from the square
to the triangular or a diluted lattice does not change the qualitative results
\cite{chang,moreira}. Elgazzar \cite{elgazzar} 
and Schulze \cite{schulze1} left the word-of-mouth limit of 
nearest-neighbour interactions and looked at longer ranges of interaction, using
a ``small world'' network \cite{elgazzar} or a power-law force \cite{schulze1}.
If the probability to convince others decays with a power of the distance, the
phase transition remains in the usual case (when a pair is needed to convince 
neighbours), but no phase transition appears in the simpler case of a single
site being able to convince \cite{schulze1}. In contrast, the Ising model may
show a transition for power-law decay of interactions even if for nearest
neighbours in one dimension no transition occurs.

Schulze also simulated a ``ghost site'' connected to all normal sites on the 
square lattice \cite{schulze2}; this ghost site convinces each normal site of 
the ghost opinion with a small probability. In marketing, this probability 
corresponds to the influence of advertising, e.g. through TV commercials.
The larger the lattice is the
smaller is the amount of advertising needed to convince the whole market.
  
A different subject is ``frustration'': What should I do if my neighbours 
to the left tell me to vote for $+1$, and those to the right tell me to vote
for $-1$ ? For the usual random sequential updating I follow first the opinion
of which I am convinced first, and later I follow the one of which I am 
convinced later: no problem. But if simultaneous updating is used, then I am
frustrated, I do not know whom to follow, and thus do not change my opinion. 
In this case, a consensus is difficult in small lattices and impossible in big 
ones \cite{ioss}. (Our introduction is mostly taken from there.) 
These difficulties are partially reduced if I am less obedient
to authority and in case of conflicting advice follow my own opinion,
defined as the majority opinion in my own past voting record \cite{Sabatelli}.
The blocking effect of frustration is also removed by a small amount of noise
\cite{Sabatelli}, when people with a low probability do not follow the
above rule. Then after sufficiently long time a nearly complete consensus
is found.

Talking about voting, the results of Brazilian elections (distribution of 
number of votes among many candidates) were reproduced quite well if the 
Sznajd model with many different opinions (instead of only $\pm 1$) is put on 
a Barab\'asi-Albert network \cite{Bernardes,ba}. If the Sznajd dynamics is
simultaneous to the growth of this network, complete consensus no longer is
possible \cite{b}. On the square lattice, if only $Q=4$ or 5 parties are 
simulated \cite{acs,b} with bounded confidence, even-odd oscillations as a 
function of the opinion number may
appear at intermediate times, with the effect that the party which was on 
second place halfway through the convicing process ends up with no votes,
just as the fourth-ranked party, while the third-ranked party at the end still
has a small number of followers. Also, bounded confidence makes it difficult 
to reach a consensus for a large number $Q$ of parties \cite{acs}. 

Returning to one dimension, Behera and Schweitzer showed that numerically their
Sznajd results cannot be distinguished from a probabilistic voter model with
interactions from nearest and next-nearest neighbours \cite{behera}.

\section{Summary}

In its first three years, the Sznajd model \cite{Sznajd}, first rejected by
Phys. Rev. Letters, found followers in four continents. Some of its results are
Ising like, others are not. More sociological numbers than only Brazilian 
elections would be nice for comparison.
  
\medskip
Thanks to F. P\"utsch and A. O. Sousa for comments on the manuscript.

\section{Appendix: Sznajd Chain and Programs}

In one dimension one can see without computer simulations why an initial 
random distribution of votes (which is not up-down periodic everywhere like an
antiferromagnet) results in a complete consensus. After some time large domains
of $+$ and $-$ are formed, with domain boundaries like $++++----$. If the 
righmost plus pair in this picture is selected, it convinces the leftmost minus 
to become plus, and the boundary is shifted to the right. If instead the 
leftmost minus pair is selected, it shifts the boundary to the left. Thus the
boundaries undergo a random walk until they annihilate each other or are 
annihilated at the two sample ends. An intermediate step of this annihilation
process are configurations like $++++-+++$ or $++-+-+--$. Here the inner 
mixed region cannot expand since in contrast to Ising models only adjacent 
pairs convince; but the inner region can shrink to annihilation when the outer
pairs are selected to convince their neighbours. Thus all boundaries finally 
vanish and we arrive at a complete consensus. These arguments apply also to 
more than two opinions, while simulations show this behaviour also in higher
dimensions, Fig.2, in not too large lattices. Simple programs for D 
and HK are listed below.
 
\begin{figure}
  \includegraphics[angle=-90,scale=0.6]{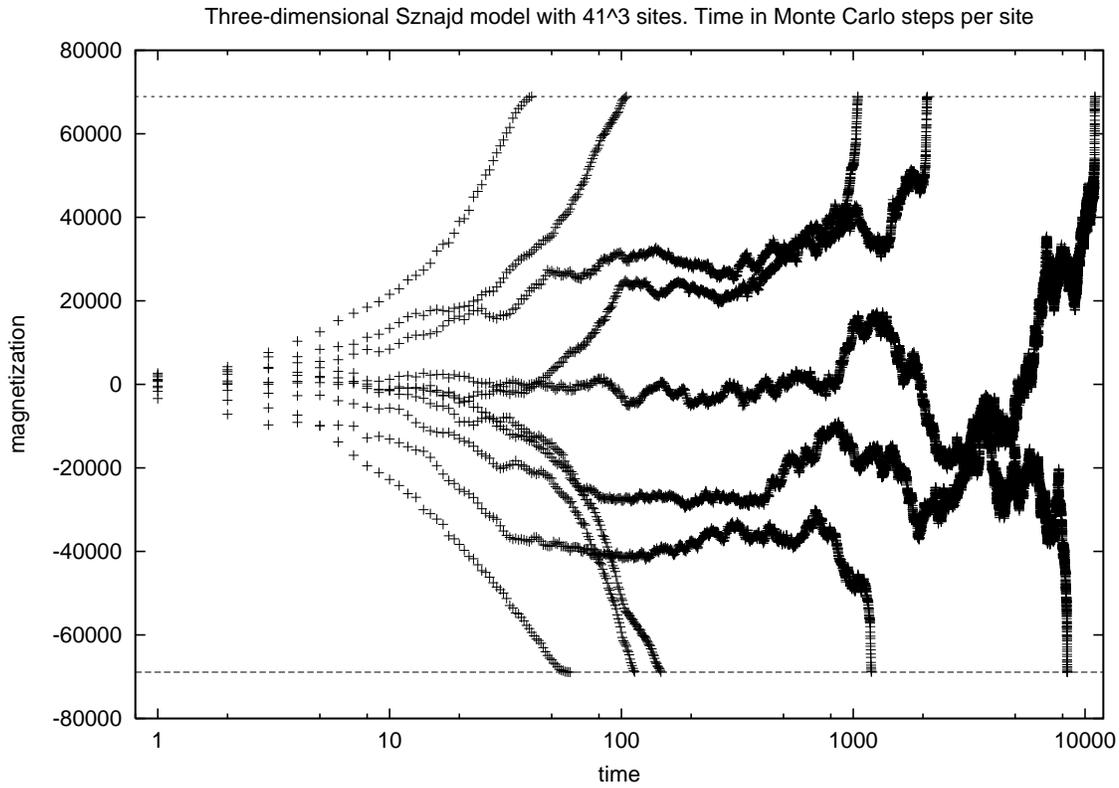}
\caption{
Variation of the magnetization, the difference between the two opinions, in ten
separate runs of a $41^3$ Sznajd lattice with neighbors in agreement convincing
their 18 neighbours. The two horizontal lines indicate complete consensus.
(For larger lattices and higher dimensions blocking cannot be excluded because
of the enormous fluctuations seen in this figure.)}
\end{figure}
\newpage
\begin{verbatim} 
      parameter(n=200,max=30,iseed=1,eps=0.4,weight=0.3)
c     Deffuant et al consensus (Weisbuch's C program)
c     s(i)=opinion of agent i; n=number of agents
      dimension s(n)
      print *, n, eps, max, iseed
      ibm=2*iseed-1
      factor=1.0d0/2147483648.0d0
      do 1 i=1,n
        ibm=ibm*16807
 1      s(i)=iabs(ibm)*factor
      do 2 iter=1,max
       do 3 i=1,n
 4      ibm=ibm*16807
        j=1+(ibm*factor+0.5)*n
        if(j.le.0.or.j.gt.n) goto 4
        if(abs(s(i)-s(j)).gt.eps) goto 3
        shift=weight*(s(j)-s(i))
        s(i)=s(i)+shift
        s(j)=s(j)-shift
 3      print *, iter, s(i)
 2    continue
      stop
      end


      parameter(n=200 ,eps=0.40,max=10,iseed=1)
c     Hegselmann-Krause consensus with sequential updating
c     s(i) = opinion of agent i, n = number of agents
      dimension s(n)
      print *, n,eps,max,iseed
      ibm=2*iseed-1
      factor=1.0d0/2147483648.0d0
      do 1 i=1,n
        ibm=ibm*16807
 1      s(i)=iabs(ibm)*factor
      do 2 iter=1,max
       do 3 i=1,n
        sum=0.0
        neighb=0
        si=s(i)
        do 4 j=1,n
         if(abs(s(j)-si).gt.eps) goto 4
         sum=sum+s(j)
         neighb=neighb+1
 4      continue
        s(i)=sum/neighb
 3      print *, iter, s(i)
 2    continue
      stop
      end
\end{verbatim}


\begin{thebibliography}{99}

\bibitem{Majorana} Majorana, E., Il valore delle leggi statistiche nella fisica
e nelle scienze sociali, Scientia 36, 58-66 (1942).

\bibitem{Schelling} Schelling, T.C., J. Mathematical Sociology 1, 143-186 
(1971). 

\bibitem{Callen} Callen E. and Shapero, D., Physics Today, July 1974, 23-28.

\bibitem{Galam} Galam, S., Gefen Y., and Shapir, Y., J. Mathematical Sociology 
9, 1-13 (1982).

\bibitem{Schweitzer} Schweitzer, F. (ed.) {\it  Self-Organization of Complex 
Structures: From Individual to Collective Dynamics}, Gordon and Breach, 
Amsterdam 1997.

\bibitem{Weidlich} W. Weidlich, {\it Sociodynamics; A Systematic Approach to 
Mathematical Modelling in the Social Sciences}. Harwood Academic Publishers, 
2000.

\bibitem{Axelrod} Axelrod, R., J. Conflict Resolut. 41, 203-226 (1997).

\bibitem{Klemm} Klemm, K., Egu\'iluz, V.M., Toral, R.  and San Miguel, S.,
preprint for Physica A (2003).

\bibitem{Deffuant} Deffuant, G., Amblard, F., Weisbuch G. and Faure, T., 
Journal of Artificial Societies and Social Simulation 5, issue 4, paper 1 
(jasss.soc.surrey.ac.uk) (2002). 

\bibitem{Hegselmann} Hegselmann R. and Krause, M., Journal of Artificial Societi
es 
and Social Simulation 5, issue 3, paper 2 (jasss.soc.surrey.ac.uk) (2002).

\bibitem{Sznajd} Sznajd-Weron K, and Sznajd, J., Int. J. Mod. Phys. C 11, 
1157-1166 (2000). 

\bibitem{Stauffer} Stauffer, D., Journal of 
Artificial Societies and Social Simulation 5, issue 1, paper 4 
(jasss.soc.surrey.ac.uk) (2002);
Bruce Schechter, New Scientist 175, \# 2357, p.42-43 (2002).

\bibitem{voter} For voter models see e.g.
Fontes, L.R., Schonmann, R.H., Sidoravicius V.,
Comm. Math. Phys. 228, 495-518 (2002)

\bibitem{Ben-Naim} Ben-Naim, E., Kaprivsky, P.L. and Redner, S., 
cond-mat/0212313.

\bibitem{galam} Galam, S., J. Stat. Phys. 61, 943-951 (1990).

\bibitem{sociocise} Stauffer, D., Computing in Science and Engineering 5, 71-75
(May/June 2003). 

\bibitem{others} Galam S. and Wonczak S., Eur. Phys. J. B 18, 183-186 (2000);
Galam, S., Zucker, J.D., Physica A 287, 644-659 (2000);
Galam, S., Chopard B., and Droz, M., Physica A 314, 256-263 (2003); 
Galam, S., Eur. Phys. J. B 25, 403-406 (2002); 
Stauffer, D., Int. J. Mod. Phys. C 13, 975-977 (2002); 
Galam, S., Physica A 320, 571-580 (2003).

\bibitem{Helbing} Helbing, D., Farkas, I. and Vicsek T., Nature 407, 487-490 
(2000).

\bibitem{Ochrombel} Ochrombel, R.. Int. J. Mod. Phys. C  12, 1091-1092 (2001)

\bibitem{Bernardes} Bernardes, A.T., Costa, U.M.S., Araujo, A.D., and 
Stauffer, D., Int. J. Mod. Phys. C  12, 159-168 (2001);  Bernardes, A.T., 
Stauffer D., and Kert\'esz, J., Eur. Phys. J. B 25, 123-127 (2002).

\bibitem{Sousa} Stauffer, D., Sousa, A.O. and Moss de Oliveira, S, 
Int. J. Mod. Phys. C  11, 1239-1245 (2000).

\bibitem{sznajdweron} Sznajd-Weron, K., Phys. Rev. E 66, 046131 (2002).

\bibitem{pmco} Stauffer, D. and de Oliveira, P.M.C., Eur. Phys. J. B 30,
587-592 (2003).

\bibitem{weron} Sznajd-Weron, K. and Weron, R., Int. J. Mod. Phys. C 13, 115
-123 (2002) and Physica A 324, 437 (2003).

\bibitem{chang} Chang, I., Int. J. Mod. Phys. C 12, 1509-1512 (2001).

\bibitem{moreira} Moreira, A.A.,  Andrade, J.S. Jr. and Stauffer, D., Int. J.
Mod.  Phys. C  12, 39-42 (2001).

\bibitem{elgazzar} Elgazzar, A.S., Int. J. Mod. Phys. C 12, 1537-1544 (2001).

\bibitem{schulze1} Schulze, C., Int. J. Mod. Phys. C 14, 95-98 (2003).

\bibitem{schulze2} Schulze, C., Physica A 324, 717-722 (2003).

\bibitem{ioss}Stauffer, D., preprint for J. Math. Sociology = cond-mat/0207598.

\bibitem{Sabatelli} Sabatelli, L. and Richmond, P.,  Int. J. Mod. Phys. C 14, 
No. 9 (2003) = cond-mat/0305015 and preprint for Physica A.

\bibitem{ba} Albert, R, Barab\'asi, A.L., Rev. Mod. Phys. 74, 47-97 (2002).

\bibitem{b} Bonnekoh, J., Int. J. Mod. Phys. C 14, No. 9 (2003) = 
cond-mat/0305125.

\bibitem{acs} Stauffer, D., Adv. Compl. Syst, 5, 97-102 (2002) and Int. J. Mod. 
Phys. C 13, 315-318 (2002).

\bibitem{behera} L. Behera and F. Schweitzer, Int. J. Mod. Phys. C 14, No. 10
(2003) = cond-mat/0306576.

\end{thebibliography}
\end{document}